\documentclass[review]{elsarticle}

\usepackage{lineno,hyperref}
\modulolinenumbers[5]

\journal{Journal of \LaTeX\ Templates}

\begin{document}

\begin{frontmatter}

\title{\boldmath Broken SU(6) symmetry and massive hybrid stars}

\author{Luiz L. Lopes\fnref{llopes@cefetmg.br}}
\address{Centro Federal de  Educa\c{c}\~ao Tecnol\'ogica de Minas Gerais Campus VIII; CEP 37.022-560, Varginha - MG - Brasil}
\ead{llopes@cefetmg.br}

\author{Debora P. Menezes\fnref{ldebora.p.m.26@gmail.com}}
\address{Universidade Federal de Santa Catarina;  C.P. 476, CEP 88.040-900, Florian\'opolis, SC, Brasil }

\begin{abstract}
In this work we revisit the quantum hadrodynamics (QHD) formalism to
investigate hyperonic and hybrid stars with hyperon-meson couplings 
fixed via broken SU(6) group, in favor of  a more general
flavor group SU(3). We also employ an additional channel, the
strangeness-hidden $\phi$ meson, which couples only to the hyperons.
In hybrid stars, the quark phase is built with the SU(3) NJL model also
with an additional vector channel in the Lagrangian.
We found that  within the models chosen the hyperon puzzle cannot be avoided by the quark-hadron
phase transition, once the hyperon threshold always happens
at lower density. Also, the contribution of the quark core in a hybrid star
is more relavant to the radius  than to the mass, and strongly depends
on the quark equation of state. We are able to reproduce 2.21
$M_\odot$ hyperonic star and 2.10 $M_\odot$ hybrid star. Both results
are in agreement with the recently
detected  hyper massive pulsar MSP J0740+6620.
\end{abstract}


\end{frontmatter}

\section{Introduction}
\label{sec:intro}

Predicted by the Soviet physicist Lev Landau even before the discovery
of the neutrons, neutron stars are the ultimate laboratory 
for cold strong interacting matter, where the density in their cores
can reach five to ten times the nuclear saturation density. 
The recent  discovery of the hyper massive  MSP J0740+6620, whose mass
range lies at $2.14^{+0.10}_{-0.09}~M_{\odot}$ with 68$\%$ credibility interval and $2.14^{+0.20}_{-0.18}~M_{\odot}$ with 95$\%$  credibility interval~\cite{Cromartie}, together with the well known  PSR J0348+0432 with mass of $2.01 \pm 0.04~M_{\odot}$~\cite{Antoniadis},
 put strong constraints in the equation of state (EoS) of dense
 $\beta$-stable matter. 

However, while the discovery of massive pulsars
points to a very stiff equation of state (EoS), the onset of hyperons
softens the very same EoS: as we  increase the density towards the
neutron star core, strange content particles -  as hyperons - 
become energetically favorable, as the Fermi  energy of the nucleons
becomes of the order of their rest masses. The appearance of hyperons 
softens the EoS and reduces the possible maximum mass of the
corresponding neutron star,  which may cause a conflict between the astrophysical
 observations and the theoretical previsions. This is known as the hyperon
 puzzle. Although some studies indicate that the hyperon threshold can be
 suppressed by either very fast rotation~\cite{Negreiros_2013} or by
 strong magnetic fields~\cite{Dex4}, this is not a consensus since a
very similar formalism does not predict such suppression~\cite{Dex6}. 
Nevertheless, these extreme conditions are not expected to be found in mostly observed
pulsars. So, in these case, as was shown in an  extensive study in
ref.~\cite{HI}, hyperons - as Thanos - are inevitable~\cite{Endgame}.

Another possibility is that the observed pulsars are indeed quark stars, composed entirely by deconfined strange matter.
If the Bodmer-Witten conjecture is true~\cite{Bodmer,Witten}, the
protons and neutrons are not the true ground state of the strong interacting matter
at high density, but rather the strange matter  is the
  preferencial ground state. If the density is high
enough the star undergoes a phase transition from hadronic to strange
matter causing all stellar matter to be converted into strange matter in a finite amount of time.

A different scenario happens if the strange matter is not the ground
state, but yet is energetically favorable at high densities. In this case the core of the neutron star undergoes a phase transition to quark matter, while the outer layers are still 
formed by hadronic matter. A star with a quark core surrounded by hadronic matter is called a hydrid star. A nice
argument  to corroborate this idea is based on the large $N_c$ expansion as shown in ref.\cite{Mc}. As the quark chemical potential exceeds the constituent quark mass,
the increase of the pressure produces a phase where chiral symmetry is
restored.  With this hypothesis, for sufficiently high densities this matter becomes strange quark matter. A very recent study~\cite{Annala}
suggests that the hybrid star is indeed the most probably
scenario for massive neutron stars. According to this work, based
  on the sound velocity analyses, if stars are very massive $M~\geq~2 M_\odot$ a hybrid star scenario is favorable, while for stars around the 
canonical mass $M~\simeq~1.4 - 1.5 M_\odot$ a pure hadronic star is more
probably.

To describe the hadronic matter we use the Quantum Hadrodynamics (QHD) formalism~\cite{Serot}. Here, besides the traditional $\sigma,~\omega$ and $\rho$ mesons~\cite{Glen},
we employ the non-standard strangeness hidden $\phi$ meson~\cite{Ellis,Lopes2013,Rafa2005,Weiss2}, which couples only to the hyperons. As pointed out 
 in our previous work~\cite{Lopes2020}, the $\phi$ meson has a crucial
 role in the  description of massive pulsars as the MSP J0740+6620.
 To fix the hyperon-meson coupling constants we break the  hybrid SU(6)  group in favor of a more general flavor SU(3) group, what allows us 
to fix all vector-meson coupling constants in terms of only one free parameter $\alpha_v$.

 In a previous work~\cite{Lopes2013}, we already studied how to use symmetry group to fix the hyperon meson coupling constants. Now we go beyond and also study quark matter in order to reproduce massive quark and hybrid stars. In the quark sector, we use the SU(3) Nambu Jona-Lasinio model NJL, which satisfies some QCD chiral symmetry aspects~\cite{Nambu,Lopes2016}.
As in the hadron phase, we employ an additional repulsive vector
channel $G_V$, to stiff the EoS. Although $G_V$ is treated as a free parameter, we 
impose some lattice QCD constraints to it~\cite{Sugano2014}. Finally,
 hybrid star matter is obtained by imposing the Maxwell construction.

section{Hadronic neutron stars}

Hadronic neutron stars -  or simply neutron stars for short - are objects composed entirely by hadrons, as neutrons and protons.
Additional degrees of freedoms as hyperons, $\Delta$'s and boson condensates can also be present~\cite{Glen}.
As the QCD, the natural tool of the standard model to describe strong interacting matter produces no results for 
dense and cold matter, we have to employ effective models. Here we use the QHD, a relativistic model which describes
the baryons as the fundamental degrees of freedom and describes their interactions via massive meson exchange.

To produce reliable neutron star properties we need to be able to reproduce realistic physical quantities that are known from 
phenomenology. There are five well known properties of symmetric
nuclear matter at the saturation point: the saturation density itself ($n_0$),
the effective nucleon mass ($M^{*}/M$), the compressibility $(K)$, the symmetry energy ($S_0$) and the binding energy per baryon ($B/A$)~\cite{Glen}.
Besides them,  the symmetry energy slope  ($L$) has attracted a lot of
attention in the last years. Althoug its true value is still a matter
of debate, most studies indicate that it has non-negligible
implications on the neutron star macroscopic
properties~\cite{Rafa,Lopes2014,Tsang,
  Micaela2017,Lattimer2014,Pais2016,Dex2019,Prov2019}. To fulfill
these constraints we use a slightly modified GM1 parametrization,
which reduces the slope from 94 MeV to 87.9 MeV. This modification also
causes a reduction in the symmetry energy from 32.5 MeV to  30.5
MeV  and makes the parametrization closer to the acceptable ones 
according to the rigid constraints imposed in \cite{Dutra2014, Micaela2017}.

The QHD Lagrangian in this work reads
(Eq. (\ref{EL1})):

\begin{eqnarray}
\mathcal{L}_{QHD} = \sum_b \bar{\psi}_b \bigg [\gamma^\mu(i\partial_\mu  - g_{b,\omega}\omega_\mu  - g_{b,\rho} \frac{1}{2}\vec{\tau} \cdot \vec{\rho}_\mu)
- (m_b - g_{b,\sigma}\sigma ) \bigg ]\psi_b     + \frac{1}{2} m_v^2 \omega_\mu \omega^\mu 
   \nonumber \\ + \frac{1}{2} m_\rho^2 \vec{\rho}_\mu \cdot \vec{\rho}^{ \; \mu}   + \frac{1}{2}(\partial_\mu \sigma \partial^\mu \sigma - m_s^2\sigma^2)  
    - U(\sigma)  - \frac{1}{4}\Omega^{\mu \nu}\Omega_{\mu \nu} -  \frac{1}{4}\bf{P}^{\mu \nu} \cdot \bf{P}_{\mu \nu}  , \label{EL1} 
\end{eqnarray}
in natural units. 
 $\psi_b$  are the baryonic  Dirac fields. Here, not only nucleons can
 be present, but we also consider the possibility of the hyperon
 presence in the neutron star core.
Because of the Pauli principle, as the number density increases, so
does the Fermi energy. Ultimately the Fermi energy of the nucleons
exceeds the mass of heavier baryons, and the conversion of some
nucleons into hyperons~\cite{Glen} becomes energetically favorable.  The $\sigma$, $\omega_\mu$ and $\vec{\rho}_\mu$ are the mesonic fields.  The $g's$
 are the Yukawa coupling constants that simulate the strong interaction,
 $m_b$ is the mass of the baryon $b$, $m_s$, $m_v$,  and $m_\rho$ are
 the masses of the $\sigma$, $\omega$, and $\rho$ mesons respectively,
 The anti-symmetric mesonic field strength tensors are given by their
 usual expressions as presented in~\cite{Glen}.

 The $U(\sigma)$ is the self-interaction term introduced in ref.~\cite{Boguta} to reproduce some of the saturation properties of the nuclear matter and is given by:
 \begin{equation}
U(\sigma) =  \frac{1}{3!}\kappa \sigma^3 + \frac{1}{4!}\lambda \sigma^{4} \label{EL2} .
\end{equation}

We also add the strangeness hidden $\phi$ meson, an additional vector
channel that couples only to the
hyperons~\cite{Ellis,Rafa2005,Lopes2020}. Therefore it does not affect any of  discussed  nuclear saturation properties.
\begin{equation}
\mathcal{L} = g_{Y,\phi}\bar{\psi}_Y(\gamma^\mu\phi_\mu)\psi_Y + \frac{1}{2}m_\phi^2\phi_\mu\phi^\mu - \frac{1}{4}\Phi^{\mu\nu}\Phi_{\mu\nu} , \label{EL3}
\end{equation}

As neutron stars  are stable macroscopic objects, we need to describe a
neutral, chemically stable  matter and hence, leptons are added as
free Fermi gases:
\begin{equation}
 \mathcal{L}_{lep} = \sum_l \bar{\psi}_l [i\gamma^\mu\partial_\mu -m_l]\psi_l , \label{EL4}
 \end{equation}
 where the sum runs over the two lightest leptons ($e$ and $\mu$).

In Tab. \ref{TL1} we display the parameters of the slightly modified GM1 model
as well as the prediction of the physical quantities and their inferred values from
phenomenology~\cite{Glen,Glen2, Dutra2014,Micaela2017,Stone}.

As we allow the hyperon threshold, we also have to fix the hyperon-meson coupling constants. Unlike the nuclear matter, we have very little experimental 
information about hyperonic matter. The main term is the hyperon potential depth fixed at the saturation density. However, just the $\Lambda$ hyperon
has the potential depth well known at -28 MeV~\cite{Glen2}. The
knowledge of the other potential depths are known with a lower degree
of precision, but widely accepted values are $U_\Sigma$ = + 30 MeV and
$U_\Xi$ = - 18 MeV ~\cite{Rafa,Weiss2}.  Unfortunately the knowledge of the hyperon potential depth is not enough to fix all  constants, once different sets of coupling constants reproduce the same potential values \cite{james_low}. Even worst is the fact that these different sets of the coupling
constants, yet predicting the same potential depth, cause  large
variations on the neutron star properties~\cite{Glen, james_high}.  So, in order to reduce the large number of free parameters, we use symmetry group theory to fix 
the coupling of the hyperons with the vector mesons and the three
potential depths to fix the coupling of the hyperons with the  scalar
meson. 

In a previous work~\cite{Lopes2020}, we used the hybrid SU(6)
group to fix all the vector mesons and a nearly SU(6)
group to fix the scalar ones. Here, we relax this condition and break
the SU(6) group in favor of the flavor SU(3) group~\cite{Lopes2013,Weiss2}. We impose that the
Lagrangian is invariant and calculate the generalized SU(3)
Clebsh-Gordon coefficients. We also consider an ideal mixing for the
$\omega$ and $\phi$ mesons (A detailed study on hyperon-meson coupling constants
within the context of symmetry groups can be found in our previous work.~\cite{Lopes2013}.
See also refs.~\cite{Swart63,Pais}). Within this approach , all the  hyperon-meson coupling constants for the vector mesons become dependent of only one parameters, $\alpha_v$.

\begin{center}
\begin{table}[ht]
\begin{center}
\begin{tabular}{|c|c||c|c|c||c|c|}
\hline 
  & Parameters & &  Phenomenology  & GM1 & Masses (MeV) \\
 \hline
 $(g_{N\sigma}/m_s)^2$ & 11.785 $fm^2$ &$n_0$ ($fm^{-3}$) & 0.148 - 0.170 & 0.153 & $M_\Lambda$ = 1116\\
 \hline
  $(g_{N\omega}/m_v)^2$ & 7.148  $fm^2$ & $M^{*}/M$ & 0.7 - 0.8 & 0.7 & $M_\Sigma$ = 1193 \\
  \hline
  $(g_{N\rho}/m_\rho)^2$ & 3.880  $fm^2$ & $K$ (MeV)& 220 - 315
                                                    &  300  & $M_\Xi$ = 1318\\
 \hline
$\kappa/M_N$ & 0.005894 & $S_0$ (MeV) & 30 - 35 &  30.5 & $m_e$ = 0.511  \\
\hline
$\lambda$ &  -0.006426 & $B/A$ (MeV) & 15.7 - 16.5 & 16.3 & $m_\mu$ = 105.6\\
\hline 
$M_N$ &  939 MeV & $L$ (MeV) & 36 - 86.8 & 87.9  & - \\ 
\hline
\end{tabular}
 
\caption{{\bf Modified} GM1 model  parameters and physical quantities inferred from experiments
  \cite{Glen2, Dutra2014, Micaela2017,Stone}. } 
\label{TL1}
\end{center}
\end{table}
\end{center}

Therefore we have for the $\omega$ meson~\cite{Lopes2013}:

\begin{equation}
\frac{g_{\Lambda\omega}}{g_{N\omega}} = \frac{4 + 2\alpha_v}{5 + 4\alpha_v}, \quad \frac{g_{\Sigma\omega}}{g_{N\omega}} = \frac{8 - 2\alpha_v}{5 + 4\alpha_v}, \quad 
\frac{g_{\Xi\omega}}{g_{N\omega}} = \frac{5 - 2\alpha_v}{5 + 4\alpha_v} .
\label{EL5}
\end{equation}

For the $\phi$ meson we have:

\begin{eqnarray}
 \frac{g_{\Lambda\phi}}{g_{N\omega}} = \sqrt{2} \bigg ( \frac{2\alpha_v - 5}{5 + 4\alpha} \bigg ) , \quad \frac{g_{\Sigma\phi}}{g_{N\omega}} = -\sqrt{2} \bigg (
 \frac{2\alpha_v + 1 }{5 + 4\alpha_v} \bigg ) \nonumber ,\\ \frac{g_{\Xi\phi}}{g_{N\omega}} =  - \sqrt{2} \bigg ( \frac{2\alpha_v + 4}{5 + 4\alpha_v} \bigg ) , \quad 
 \frac{g_{N\phi}}{g_{N\omega}} = 0 \label{EL6}.
\end{eqnarray}

And finally for the $\rho$ meson:

\begin{equation}
 \frac{g_{\Sigma\rho}}{g_{N\rho}} = 2\alpha_v, \quad \frac{g_{\Xi\rho}}{g_{N\rho}} = 
 2\alpha_v - 1 , \quad \frac{g_{\Lambda\rho}}{g_{N\rho}} = 0. \label{EL7}
\end{equation}
To solve the equations of motion, we use the mean field approximation (MFA), where the meson fields are replaced by their expectation values, i.e:  $\sigma$ $\to$ $\left < \sigma \right >$ = $\sigma_0$,   $\omega^\mu$ $\to$ $\delta_{0 \mu}\left <\omega^\mu  \right >$ = $\omega_{0}$, $\phi^\mu$ $\to$ $\delta_{0 \mu}\left <\phi^\mu  \right >$ = $\phi_{0}$  and   $\rho^\mu$ $\to$ $\delta_{0 \mu}\left <\rho^\mu  \right >$ = $\rho_{0}$.
Applying the Euler-Lagrange formalism to the sum of  Eqs.~(\ref{EL1}) and 
(\ref{EL3}) we obtain  the following equation of motion:

\begin{equation}
[\gamma_0(i\partial^0 - g_{b, \omega}\omega_0 - g_{b, \phi}\phi_0 - g_{b, \rho}\rho_0) - i\gamma_j\partial^j - M^{*}_b]\psi_b = 0, \label{EL8}
\end{equation}
where we define $M^{*}_b~\doteq~m_b - g_{b,\sigma}\sigma_0$ as the effective baryon mass.  Using the quantization rules ($E = i\partial^0$, $i\partial^j = k$) we easily obtain the eigenvalue for the
energy:  

\begin{equation}
E_b = \sqrt{k^2 + M^{*2}_b} + g_{b,\omega}\omega_0 + g_{b,\phi} + g_{b,\rho} \frac{I_{3b}}{2} \rho_0 ,  \label{EL9}
\end{equation}
where $I_{3b}$ is the isospin projector, and assumes the value of +1 for p, $\Sigma^+$ and $\Xi^0$,  zero for $\Lambda^0$ and $\Sigma^0$ and -1 for n, $\Sigma^-$, and $\Xi^-$.

When we set $\alpha_v$ = 1, we recover the SU(6) parametrization
for the vector mesons. In this case, the $\omega$ meson couples to
hypercharge, while the $\rho$ meson couples to isospin, as proposed
by Sakurai~\cite{Sakurai60}.

Leptons are added as a free Fermi gas, with energy:
\begin{equation}
E_l = \sqrt{k^2 + m^{2}_l}. \label{EL10}
\end{equation}

To construct the equation of state (EoS) for this many body system of
leptons and strongly interacting baryons we use the Fermi-Dirac statistics.
 As the thermal energy of a stable neutron star is much lower
  than the Fermi energy of its particles, $T= 0$ is a good
  approximation.  For the baryons  the solution for the energy density is straightforward~\cite{Greiner2}:

\begin{equation}
\epsilon = \frac{1}{\pi^2}\int_0^{k_f} \sqrt{k^2 + M^{*}_b}k^2 dk , \label{EL11}
\end{equation}
the same expression is valid for the leptons, we just have to replace the effective baryon mass $M^*_b$ to the lepton mass $m_l$.

In MFA the contribution of the mesonic fields to the energy density  is given by~\cite{Glen,Lopes2012,Lopes2013}
\begin{equation}
\epsilon_m =  \frac{1}{2}\bigg ( m_s^2\sigma_0^2 + m_v^2\omega_0^2  + m_\rho^2\rho_0^2 \bigg ) + U(\sigma) , \label{EL12}
\end{equation}

The total energy density is the sum of the energy density of all fields (baryons, leptons and mesons).
 Finally the expected values of the mesonic fields are calculated
either from the Euler-Lagrange equations or by imposing that the total energy density be stationary at fixed baryon density~\cite{Glen}:

\begin{equation}
\bigg ( \frac{\partial \epsilon}{\partial \sigma_0} \bigg ) =  \bigg ( \frac{\partial \epsilon}{\partial \omega_0} \bigg)
= \bigg ( \frac{\partial \epsilon}{\partial \rho_0} \bigg ) = \bigg ( \frac{\partial \epsilon}{\partial \phi_0} \bigg ) = 0 . \label{EL13}
\end{equation}

To calculate every particle population at a fixed density we impose
electric charge neutrality and chemical equilibrium:

\begin{equation}
\mu_{bi} = \mu_n -e_{bi}\mu_e , \mu_\mu =  \mu_e ; \quad \sum_f e_fn_f = 0 , \label{EL14}
\end{equation} 
where $\mu_{bi}$ and $e_{bi}$  are the chemical potential and electric
charge of the i-th baryon respectively. At zero temperature,
the chemical potentials coincide with the energy eigenvalues given in Eqs.~(\ref{EL9}) and~(\ref{EL10}); $\mu_e$ and $\mu_\mu$ are the electron
and muon chemical potential respectively; $n$ is the number density
and the sum in $f$ runs over all the  fermions.

To obtain the complete  EoS we calculate the pressure via thermodynamics:
\begin{equation}
p = \sum_f \mu_f n_f - \epsilon , \label{EL15}
\end{equation}

\begin{figure}[t]
\begin{tabular}{cc}
\centering 
\includegraphics[scale=.51, angle=270]{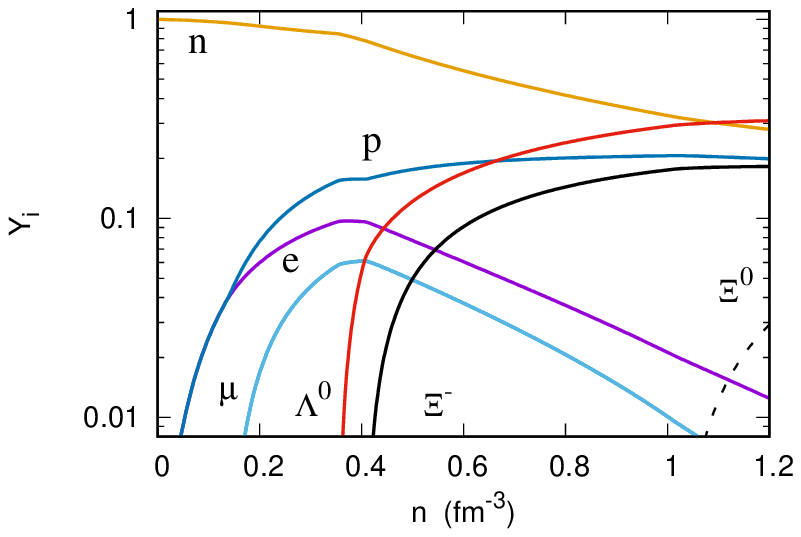} &
\includegraphics[scale=.51, angle=270]{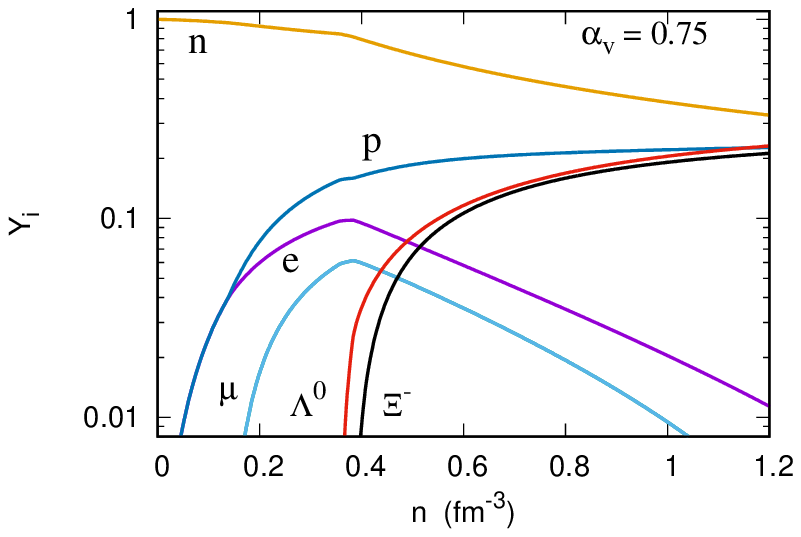}\\
\includegraphics[scale=.51, angle=270]{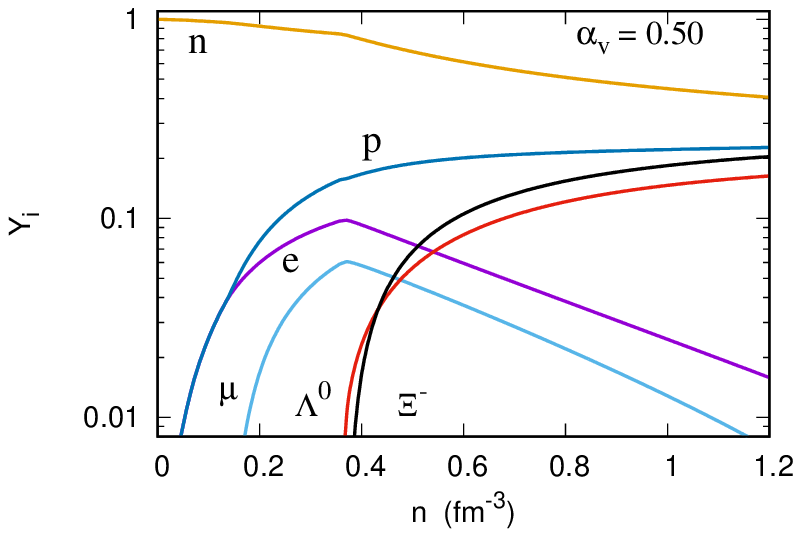} &
\includegraphics[scale=.51, angle=270]{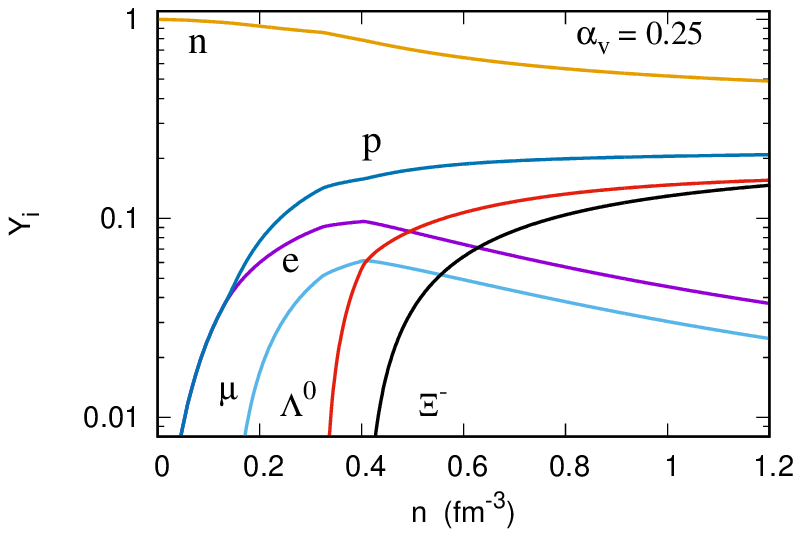}\\
\end{tabular}
\caption{(Color online) Particle population for the SU(6) group and
  different values of $\alpha_v$. } \label{FL1}
\end{figure}

\subsection{Results}

We next choose some values to $\alpha_v$ and calculate the vector meson coupling constants.  In our previous work~\cite{Lopes2013}, we also fixed the scalar meson within symmetry group context. However, although the symmetry groups reproduce good results for $\alpha_v$ close to 1.0, as we move away from this value, the hyperon potential depths become unrealistic. So, here the scalar meson coupling constants are fixed in order to reproduce the more acceptable values of the potential depth:$U_\Lambda$  = $-28$ MeV,  $U_\Sigma=$  = $+30$~MeV and $U_\Xi$  = $-18$~MeV.  The calculated coupling constants  are displayed in Tab.~\ref{TL2}.

\begin{table}[ht]
\begin{center}
\begin{tabular}{|c|c|c|c|c|}
\hline
 - & $\alpha_v = 1.00$ & $\alpha_v = 0.75$ &  $\alpha_v = 0.50$ &  $\alpha_v = 0.25$   \\
\hline
 $g_{\Lambda\omega}/g_{N\omega}$        & 0.667 & 0.687   & 0.714 & 0.75   \\
\hline
 $g_{\Sigma\omega}/g_{N\omega}$         & 0.667 & 0.812  & 1.00 & 1.25   \\
 \hline
$g_{\Xi\omega}/g_{N\omega}$           & 0.333 & 0.437  & 0.571 & 0.75   \\
\hline
$g_{\Lambda\phi}/g_{N\omega}$           & -0.471 & -0.619  & -0.808 & -1.06   \\
\hline
$g_{\Sigma\phi}/g_{N\omega}$           & -0.471 & -0.441  & -0.404 & -0.354   \\
\hline
$g_{\Xi\phi}/g_{N\omega}$           & -0.943 & -0.972  & -1.01 & -1.06   \\
\hline
$g_{\Sigma\rho}/g_{N\rho}$           & 2.0 & 1.5  & 1.0 & 0.5   \\
\hline
$g_{\Xi\rho}/g_{N\rho}$           & 1.0 & 0.5  & 0.0 & -0.5   \\
\hline
$g_{\Lambda\sigma}/g_{N\sigma}$           & 0.610 & 0.626  & 0.653 & 0.729   \\
\hline
$g_{\Sigma\sigma}/g_{N\sigma}$           & 0.403 & 0.514  & 0.658 & 0.850   \\
\hline
$g_{\Xi\sigma}/g_{N\sigma}$           & 0.318 & 0.398 & 0.500 & 0.638   \\
\hline
\end{tabular}
 \caption{Hyperon-meson coupling constants for different values of $\alpha_v$. When we impose $\alpha_v$ = 1 we recover the hybrid group SU(6).}\label{TL2}
 \end{center}
 \end{table}

We plot in Fig.~(\ref{FL1}) the particle population for different
values of $\alpha_v$. We see that  the hyperon population depends strongly
on the hyperon-meson coupling constants. As we move away
from the SU(6) ($\alpha_v$ = 1) towards $\alpha_v$ = 0.25, we see a suppression of the strangeness content particles. 
Indeed, only when $\alpha_v$ = 1 we have the presence of $\Xi^0$
hyperon and only in this case, a strangeness content particle is the
most populated particle - the $\Lambda^0$ - at very high densities (n
$>~1~fm^{-3}$). For $\alpha_v$ = 0.75 the $\Lambda^0$ and the $\Xi^-$
population are of the same order as the proton one.
Reducing the value of $\alpha_v$ makes the neutron and the proton  the
most populated particles for all densities. The suppression of the
$\Xi^-$ makes the lepton fraction to increase for low values of
$\alpha_v$. For instance, the muon population at 1 $fm^{-3}$ is about
30 times higher in $\alpha_v$ =0.25 as compared with the SU(6) case. A
curious case is $\alpha_v$ = 0.50, being the only one which produces
the $\Xi^-$ instead of the $\Lambda^0$ as the most populated hyperon at high densities.

\begin{figure}[htb] 
\begin{centering}
 \includegraphics[angle=270,
width=0.74\textwidth]{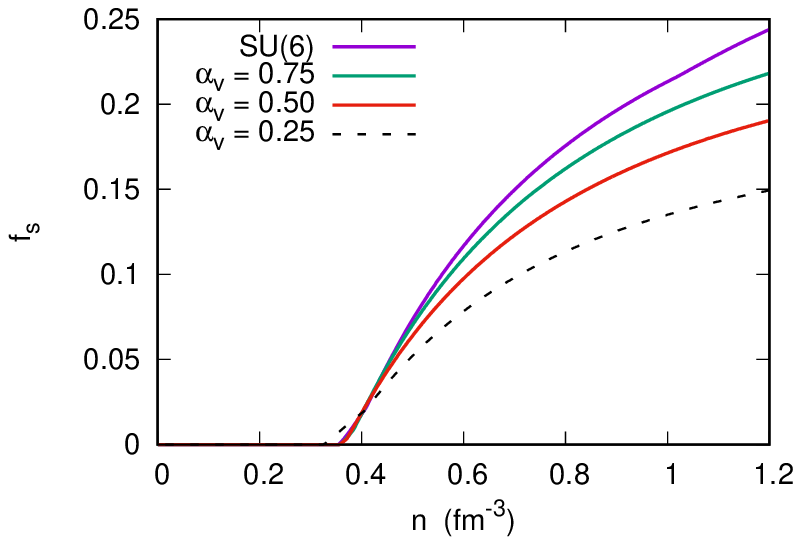}
\caption{(Color online) Strangeness fraction as a function of $\alpha_v$.} \label{FL2}
\end{centering}
\end{figure}

A more clever way to understand the suppression of strangeness content particle is, instead of looking at the individual hyperon population, look at the strangeness fraction, $f_s$, defined as:

\begin{equation}
 f_s = \frac{1}{3} \frac{n_i |s_i|}{n} , \label{EL16}
\end{equation}
where $s_i$ is the strangeness of the $i-th$ baryon. The results are plotted in Fig.~\ref{FL2}.

\begin{figure*}[htb]
\begin{tabular}{cc}
\includegraphics[width=5.6cm,height=7.0cm,angle=270]{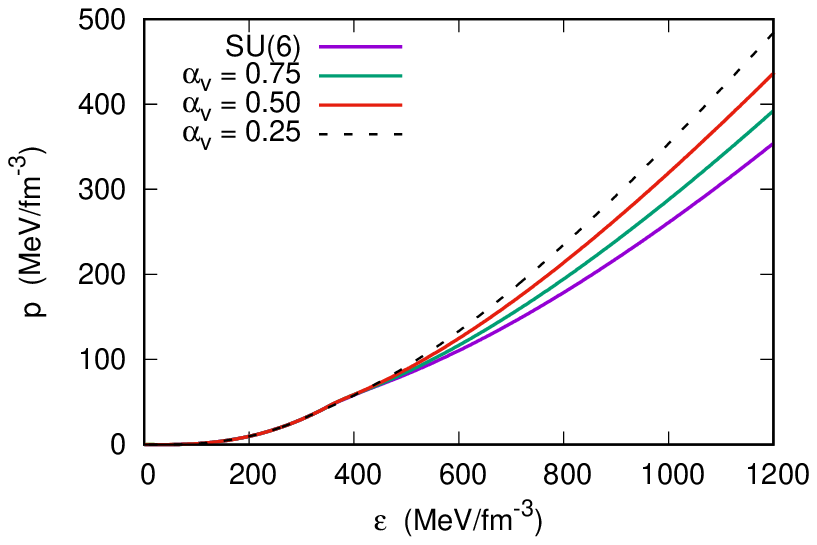} &
\includegraphics[width=5.6cm,height=7.0cm,angle=270]{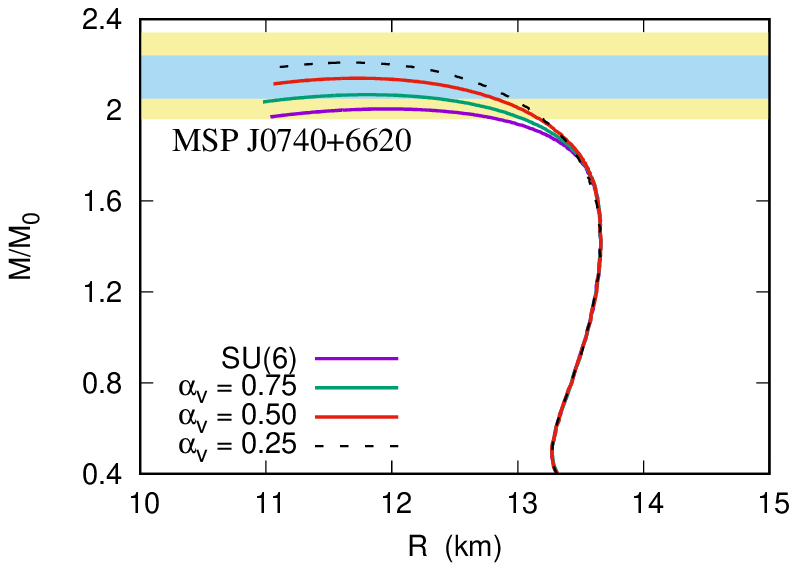} \\
\end{tabular}
\caption{(Color online) (right) EoS and (left) mass-radius relation
  for a neutron star for different values of $\alpha_v$.
  The hatched areas correspond to 68$\%$  and 95$\%$ credibility interval for the MSP J0740+6620. } \label{FL3}
\end{figure*}

As we can see, there is a direct relation between $\alpha_v$ and the strangeness fraction, as also pointed in ref.~\cite{Lopes2013}. When we move away from SU(6) we increase the repulsion of the hyperons, by increasing the $Y-\omega$ and the $Y-\phi$ 
coupling constants as shown in Tab.~\ref{TL2}. This reduces the hyperon population at high densities, reducing the strangeness fraction.

In Fig.~\ref{FL3} we plot the EoS and the respective mass-radius relation by solving the TOV equations~\cite{TOV} for the discussed values of $\alpha_v$. As expected, there is a clear relation between the $\alpha_v$ and the maximum mass neutron star.
As we reduce the value of $\alpha_v$, we increase the value of $\omega$ and $\phi$ fields, increasing the hyperon repulsion. This makes the EoS stiffer, which in turn
increases the maximum mass. As the EoS needs to be constrained by the massive known pulsars~\cite{Antoniadis,Cromartie}, we use the MSP J0740+6620 mass range of 
$2.14^{+0.10}_{-0.09}~M_{\odot}$ at 68$\%$ credibility interval (ligth blue) and $2.14^{+0.20}_{-0.18}~M_{\odot}$ at 95$\%$  credibility interval (light yellow) as
error bars. As we can see all parametrizations used in this work agree
with the 95$\%$  credibility interval, while  just the SU(6) lies
  out of  the range of the  68$\%$  credibility interval. Our study indicates that massive neutron stars with hyperons are not ruled out. The main neutron star properties are shown in Tab.~\ref{TL3}.

\begin{table}[ht]
\begin{center}
\begin{tabular}{|c|c|c|c|c|}
\hline
 $\alpha_v$ & $M_{max}/M_\odot$ & $ R (km)$ & $n_c$ ($fm^{-3}$) & $f_{sc}$    \\
\hline
 SU(6)                     & 2.00 & 11.95   & 0.92 & 0.200   \\
\hline
 $\alpha_v$ = 0.75         & 2.07 & 11.80  & 0.93 & 0.185   \\
 \hline
$\alpha_v$ = 0.50          & 2.14 & 11.73  & 0.94 & 0.164   \\
\hline
$\alpha_v$ = 0.25          & 2.21 & 11.67  & 0.93 & 0.129   \\
\hline

\end{tabular}
 \caption{Neutron stars main properties for different values of
   $\alpha_v$; the subscript $c$ means central density.}\label{TL3}
   \end{center}
 \end{table}

 The maximum obtained mass varies from 2.00$M_{\odot}$ for $\alpha_v$ = 1 to 2.21$M_{\odot}$ for $\alpha_v$ = 0.25. We see that while the strangeness fraction drops from $20\%$ to $12.9\%$ for the maximum mass, the central density $n_c$ hardly changes.
Another important point concerns  the radii of the canonical  1.4$M_\odot$ stars.
Although some studies pointed out that the radii of the canonical stars
could be as larger as 17 km \cite{17km}, nowadays this value is
believed to be significant lower. Conservative results point towards a
maximum radius of 13.9 km~\cite{Hebeler}, yet, more radical studies point to 13 km as the maximum radius \cite{Lattimer2013,Lattimer2014}.
 Studying the deformability parameter of the canonical star, the LIGO and Virgo collaboration stated that its value lies in the range $70 \leq \Lambda_{1.4} \leq 580$ \cite{PRL121} and this restriction imposed another constraint to the radius of the corresponding star. 
According to \cite{malik}, the values should lie in the region $11.82$ km $ \leq R_{1.4 M_{\odot}} \leq 13.72$ km and
according to \cite{PRL121}, in the range $10.5$ km $ \leq R_{1.4
  M_{\odot}} \leq 13.4$ km. As for all our parametrizations we have
13.68 km for  1.4$M_\odot$ star (and no hyperon at such mass value),
whichever constraint we consider correct, we see that our results for
the radii are very close to the border of these ranges. Nevertheless, a very new result  
indicates that the canonical neutron star radius cannot exceed 11.9
km~\cite{Capano2020}. This radius value together with the mass of the MSP J0740+6620
may indicate that a profound revision either on the nuclear theory or
the general relativity \cite{clesio2019} may be needed. For instance, in
ref.~\cite{Lopes2018}, we have shown that the onset of a
new, yet unknown free parameter, can produce massive and compact
neutron star families. However, this is not a closed subject, once recent
results coming from NICER experiments~\cite{NICER1,NICER2} as well 
neutron skins~\cite{PRL120} point towards maximum radius values 
of 13.85 km, 14.26 km and 13.76 km respectively. We reinforce here
that our result of 13.68 km agrees with ref.~\cite{NICER1,NICER2,PRL120}
besides ref.~\cite{Hebeler,malik}.

\section{Quark stars}

In nature, deconfined quark matter certainly existed in the early
universe when the temperature was very high. Up today, it is not clear if deconfined quark matter
exists in the core of massive neutron stars. Moreover, if the Bodmer-Witten conjecture is
true~\cite{Bodmer,Witten}, pulsars with central densities above a certain limit
should be converted into  strange stars. The main difference between
quark stars - or strange stars - and the conventional neutron stars
composed of baryons, is the fact that while the neutron star is
bounded by gravity, strange stars are bounded by the strong force itself.

To describe a quark star, we need a quark matter EoS. As in the hadronic case, the natural tool to describe these 
matter is the QCD. And as we did before, we resort to an effective model.
In the limit of vanishing quark masses, we expect the QCD to present chiral symmetry. Long before the QCD
was known to be the theory of strong interactions, phenomenological indications for the existence of chiral symmetry 
came from the study of the nuclear $\beta$ decay. An effective model that is well known to present these features is the Nambu Jona-Lasinio model~\cite{Nambu}.
Here we use its SU(3) version, whose Lagrangian includes a scalar, a pseudo-scalar and
the t'Hooft six-fermion interaction - needed to model the axial
symmetry breaking~\cite{Kun1,Kun2,DebPRC2014} and reads:

\begin{eqnarray}
\mathcal{L}_{NJL} =  \bar{\psi}_f  [\gamma^\mu(i\partial_\mu -  m_f]\psi_f   + \nonumber \\
+ G_s \sum_{a=0}^8[(\bar{\psi}\lambda_a\psi)^2 + (\bar{\psi}\gamma_5\lambda_a\psi)^2] 
- K\{det[\bar{\psi}(1 + \gamma_5)\psi + det[\bar{\psi}(1 -\gamma_5)\psi]\} \label{EL17}
\end{eqnarray}
where $\psi_f$ are the quark Dirac fields, with three flavors, $m_f$ = diag$(m_u, m_d, m_s)$ are the
current quark masses, $\lambda_a$ are the eight Gell-Mann flavor matrices and $G_s$ and $K$ are dimensionful coupling constants.
Unlike the QHD model for baryons,  where the interaction is mediated by massive mesons,  the NJL model has no mediator, and the interaction is a direct quark-quark point-like scheme (see ref.~\cite{Kun2} to see the Feynman diagrams).
This makes the NJL  a non-renormalizable model, and a cutoff is needed
to obtain physical results. The SU(3) NJL is adjusted according to five main physical parameters:
the $\pi$, $\eta$ and $\sigma$ meson masses, as well as the pion and
$\eta$ decay coupling constants, $f_\pi$ and $f_\eta$. The parameters
we choose to use (HK), the physical predictions and experimental values are given in Tab. \ref{TL4}.

\begin{center}
\begin{table}[ht]
\begin{center}
\begin{tabular}{|c|c||c|c|c||c|}
\hline 
  & Parameters & &  Phenomenology  & SU(3) NLJ \\
 \hline
 $m_u = m_d$ & 5.5 MeV  & $m_\pi$ (MeV) & 128 -138 & 138 \\
 \hline
  $m_s$ & 135.7 MeV   & $m_\eta$ (MeV) & 487  & 549  \\
  \hline
  $\Lambda$ & 631.4 MeV   & $m_\sigma$ (MeV) & 668 &  700  \\
 \hline
$G_s\Lambda^2$ & 1.835 & $f_\pi$ (MeV) & 93 &  93   \\
\hline
$K\Lambda^5$ &  9.29 & $f_\eta$ (MeV) & 94.3 & 84 - 102 \\
\hline 
\end{tabular}
 
\caption{SU(3) NJL  parameters and physical quantities infered from experiments
\cite{Kun2}. } 
\label{TL4}
\end{center}
\end{table}
\end{center}

Now, as in the hadronic case, we add an additional vector channel.
Here we use an universal vector coupling:

\begin{equation}
\mathcal{L}_{NJLv}= -G_V(\bar{\psi}\gamma^\mu\psi)^2 .\label{EL18}
\end{equation}

In the
phase diagram, the vector term weakens and delays the phase transition of the
chiral restoration, and can potentially alter the nature from
chiral transition  to the  color-superconducting (CSC)
phase~\cite{kit2002}. The mathematical formalism of the
vector term shows that it acts similarly to the $\omega$ meson in
 QHD  models, creating an additional 
repulsion between the quarks and stiffens the EoS \cite{klahn}. This
effect is desirable once we need to construct an EoS stiff enough to
simulate the two solar mass MSP J074+6620 pulsar.

Assuming the mean field approximation (MFA) we can rewrite the quark-quark interaction in terms of the scalar condensates and the quark number density:~\cite{MB}

\begin{eqnarray}
(\bar{\psi}\psi)^2 = 
2\langle\bar{\psi}\psi\rangle (\bar{\psi}\psi) - \langle\bar{\psi}\psi\rangle^2 ,\nonumber \\
(\bar{\psi}\gamma^0\psi)^2 = 
2\langle\bar{\psi}\gamma^0\psi\rangle (\bar{\psi}\gamma^0\psi) - \langle\bar{\psi}\gamma^0\psi\rangle^2. \label{EL19}
\end{eqnarray}

The dressed quark masses, $M_f$
are determined by a coupled set of gap equations~\cite{DebPRC2014},
where $\phi_f$ is the scalar quark condensed of flavor $f$: $\phi_f
= \langle\bar{\psi}_f\psi_f\rangle$: 

\begin{equation}
 M_f =  m_f - 4G_s\phi_f +2K\phi_i\phi_j \label{EL20}
\end{equation}
with $i~\neq~j~\neq~f$ and:

\begin{equation}
 \phi_f =  \frac{-\nu M_f}{2\pi^2} \int_{k_{Ff}}^\Lambda \frac{k^2d k}{\sqrt{M_f^2 + k^2}},\label{EL21}.
\end{equation}
where $k_{Ff}$ is the Fermi momentum of the quark f and $\nu = 6$ is
the color-spin degeneracy factor. In the same way, the vector channel
induces a displacement of the energy eigenvalue (as well as in the
chemical potential at  $T=0$), i.e.,

\begin{equation}
 E_f = \mu_f =  \sqrt{M_f^2 + k_{Ff}^2} + 2G_V n, \label{EL22}
\end{equation}
where $n$ is the total quark number density: $n = \sum_f n_f$. Now the
energy density is obtained by taking into account the vacuum and the in-medium contributions.
We can write~\cite{DebPRC2014,Hana2001}

\begin{eqnarray}
 \epsilon = \sum_f \bigg [\frac{\nu}{2\pi^2} \int_{k_{Ff}}^\Lambda \sqrt{M_f^2 + k^2} k^2 dk + 2G_s\phi_f^2 \bigg ] \nonumber 
 +G_Vn^2 + 4K\phi_i\phi_j\phi_k - \epsilon_{vac} , \label{EL23}
\end{eqnarray}
where the constant $\epsilon_{vac}$ is the vacuum energy, introduced
in order to set the energy density of the physical vacuum $(k_F = 0)$
equal to zero. It is also worth mentioning that Eq.~(\ref{EL18}) is
not the only way to introduce a vector channel in the NJL models (see ref.~\cite{DebPRC2014} for more details). 

As in the hadronic case, leptons are added as a free Fermi gas, as
required by  charge neutrality and chemical stability.   The relations
between the chemical potentials and the number density of different
particles are given by~\cite{Rhabi}:

  \begin{eqnarray}
  \mu_s =\mu_d = \mu_u + \mu_e , \quad \mbox{and} \quad \mu_e = \mu_\mu , \nonumber \\
  n_s + n_\mu = \frac{1}{3}(2n_u -n_d - n_s). \label{NEQ}
  \end{eqnarray}

The energy density is obtained
via the thermodynamic relation given in (Eq.~\ref{EL15}).

In order to obtain physical results we need to fix the $G_v$ coupling constant.
 While in  QHD the non standard vector channel $\phi$ introduces
no new free parameters because all hyperon-vector mesons can be fixed throughout symmetry group arguments, unfortunately,  in the NJL,  this is not the case. In
most works the $G_v$ is treated  just as a free parameter
~\cite{DebPRC2014,MB,DebJCAP2019,Hana2001,Shao}.
Nevertheless,  in other works, the autors have tried to fix the
$G_v$ coupling  from direct comparisons with the lattice QCD
(LQCD) results. 

In ref.~\cite{kit2002} studying the interplay
between chiral transition and CSC phase, the authors fixed
$G_V$ in the range $0.2G_s< G_v <0.3G_s$ in order to reproduce the
LQCD; in ref.~\cite{Contrera2014}  $G_v$ was fixed in the range $0.283G_s<$ $G_v <0.373G_s$
in order to reproduce the slope of the pseudo-critical temperature for the chiral phase transition at low chemical potential 
extracted from LQCD simulations; also to reproduce the pseudo-critical
temperature, in ref.~\cite{Hell2011} $G_V$ was found to be in the range $0.25G_s<$ $G_V <0.4G_s$ and finally, in ref.~\cite{Sugano2014}  a very restrictive choice was
made and the $G_V$ = $0.33G_s$. We use this value,  $G_V$ = $0.33G_s$
as a limit of acceptable values that agree with the LQCD. We use then four different parametrizations for $G_V$: $G_V = 0.00$, $G_V = 0.11G_s$, $G_V = 0.22G_s$, and $G_V = 0.33G_s$. However, as the value $G_V = 1.0G_s$  up today can be found in the literature~\cite{DebPRC2014,DebJCAP2019,Hana2001}, we also use this value for matter of completeness and comparison. Nevertheless we have to keep in mind that such value is away above what is expected from LQCD.

\subsection{Results}

\begin{figure}[htb] 
\begin{centering}
 \includegraphics[angle=270,
width=0.74\textwidth]{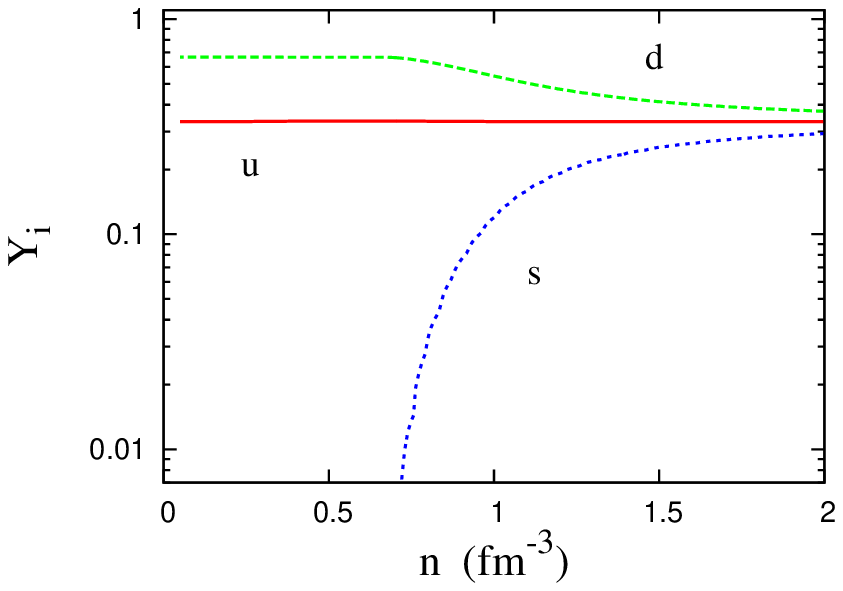}
\caption{(Color online) The quark population for all values of $G_V$. Unlike in hadronic phase, the lepton population is insignificant for all densities. } \label{FL4}
\end{centering}
\end{figure}

In  Fig.~\ref{FL4} we plot the  particle population for all values of $G_V$.
Unlike the baryon population, the quark population  does not
  depend  on  the coupling constant because, as shown in
  Eq.~(\ref{EL22}), the displacement in the chemical potential is the
  same for all quarks and for all values of $G_V$. As  $G_s$ and $K$
  coupling constants are equal to all quarks, the main  difference is
  generated by the mass of the $s$ quark. With our choice following
ref.~\cite{Kun1}, the $s$ quark onset happens around 0.66
$fm^{-3}$. Also, the lepton population is insignificant through out
the star. The electron population  has a maximum of  only $Y_e =0.002$
next to the $s$ quark threshold, then it drops for higher
densities. The muons are absent due the fact that their mass is 
bigger than the mass of the light quarks.

\begin{figure*}[ht]
\begin{tabular}{cc}
\includegraphics[width=5.6cm,height=7.0cm,angle=270]{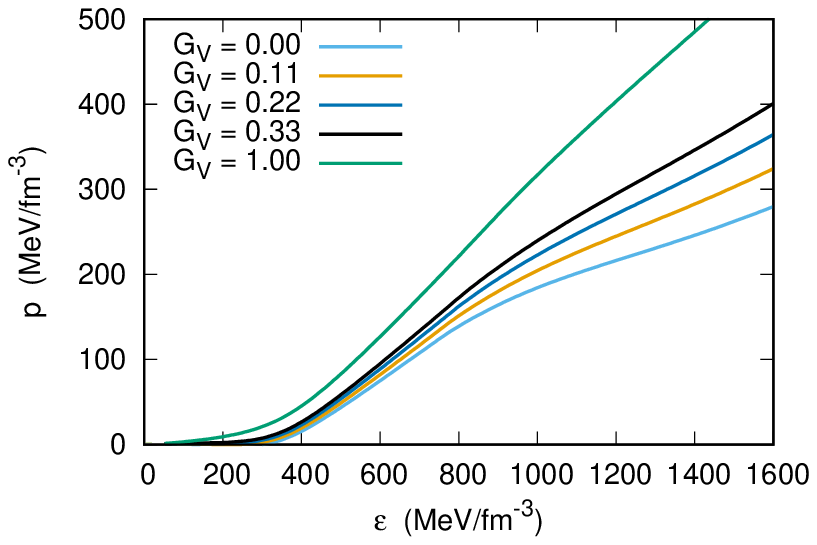} &
\includegraphics[width=5.6cm,height=7.0cm,angle=270]{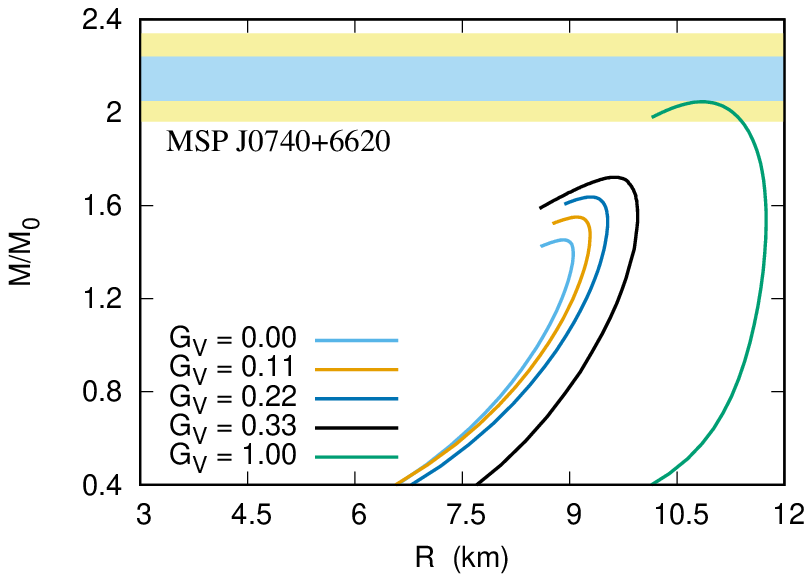} \\
\end{tabular}
\caption{(Color online) (right) EoS and (left) mass-radius relation
  for a quark star for different values of $G_V$. Strength is the vector coupling, stiff is the EoS, as well higher is the maximum mass.} \label{FL5}
\end{figure*}

  As for the EoS itself, it is plotted in Fig.~\ref{FL5} alongside
some of the corresponding macroscopic properties of the quark stars, i.e.,
the mass-radius relation for the different values of $G_V$. As expected, there is
a simple relation between the strength of the $G_V$ and the EoS, which reflects 
in the maximum mass. Increasing the $G_V$, we stiff the EoS and
increase the maximum mass. For realistic values of $G_V$, our maximum
mass is quite below the experimental limit of the MSP J0740+6620. But,
if we increase the value of $G_V$ we are able to reproduce a 2.05
$M_\odot$ quark star. Nevertheless we have to keep in mind that $G_V =
1.0G_s$ is way above the values expected from
LQCD~\cite{Sugano2014}. Therefore, instead of considering artificial
and unphysical values of $G_V$,  we accept the fact that massive
neutron stars can hardly be described as quark stars. The main results are resumed in Tab.~\ref{TL5}.

\begin{table}[ht]
\begin{center}
\begin{tabular}{|c|c|c|c|c|}
\hline
 $G_V$ & $M/M_\odot$ & $ R (km)$ & $n_c$ ($fm^{-3}$) & $f_{sc}$    \\
\hline
 $0.00G_s$         & 1.46 & 8.92   & 1.17 & 0.186   \\
\hline
 $0.11G_s$         & 1.55 & 9.12  & 1.16 & 0.184   \\
 \hline
$0.22G_s$          & 1.64 & 9.31  & 1.14 & 0.177   \\
\hline
$0.33G_s$          & 1.73 & 9.59  & 1.10 & 0.165   \\
\hline
$1.00G_s$          & 2.05 & 10.90  & 0.87 & 0.055  \\
\hline

\end{tabular}
 \caption{Quark star main properties for different values of $G_V$.}\label{TL5}
 \end{center}
 \end{table}

For realistic values of $G_V$, the maximum quark star masses varies
from 1.46 $M_\odot$ to 1.73 $M_\odot$. We also see that the higher the
$G_V$ value, the higher the radius of the maximum mass quark star. 
In general, quark stars are denser than hadronic ones.
On the other hand, increasing $G_V$ causes a reduction of 
the central baryon number $n_c$. In the same way, it causes a
reduction of the strangeness fraction. For $G_V =1.00G_s$ the
strangeness fraction is only $5.5\%$. Such low value is  below the
strangeness fraction found in hadronic stars, even  with a strong hyperon-hyperon repulsion.

Also pointed out in ref.~\cite{DebJCAP2019} and other references therein,
the hadronic neutron star could be a meta-stable system, which
  eventually collapses. If the original neutron star
had a mass beyond 1.73 $M_\odot$ (the higher mass value for a realistic vector channel)
it would become a black hole. 
However, a lower mass neutron star could become a quark star. The
  other possibility is that the metastable hadronic star can face a
  transition to a hybrid star \cite{Melrose}, an object with both, hadron
  and quark matter \cite{hybrid}, as  discussed in the next section.
  
  \section{Hybrid Stars}

As we said earlier, it is not clear if deconfined quarks are present in the core of massive pulsars. However, from a phenomenological point of the view, as  the density increases towards the star core, quarks can become more energetically favorable than baryons, and ultimately  the neutron star core may be composed of deconfined quarks. If the entire star does not convert itself into a quark  star as suggested by the Bodmer-Witten conjecture, the final
composition is a quark core surrounded by a hadronic  layer. This is
what is generally called a hybrid star. Nevertheless the physics and
formalism used to construct a hybrid star still present some bias. 

Some authors ~\cite{Beth,Serot2} suggest that 
the Maxwell construction is more suitable to build a hybrid star.
Within Maxwell constructions the quark-hadron phase transition happens
at constant pressure. However, it implies that the quark and the hadron phases  are
spatially  separated and there is a discontinuity in the electron
chemical potential, although the neutron chemical potential is continuous.

On the other hand ref.~\cite{Glen} argues that the Gibbs condition is
better as  Maxwell construction can be  $\beta$ unstable at the interface between the phases. In Gibbs condition the pressure and all chemical
  potentials (including the electron chemical potential) are
  continuous. 
  Under these construction, instead of spatially separated phases,  quarks and hadrons coexist in a mixed state, generating one intermediate phase in between the hadronic and the quark phases.

  The differences between Maxwell and Gibbs construction was already checked in ref.~\cite{Chiba1,Chiba2,Paoli}.  The authors    performed studies on hybrid stars
with both  constructions. They all concluded that
there is no significant difference on the macroscopic properties of the
hybrid stars. Due to this fact, in this work we use a Maxwell
construction, which is simpler. In this case we impose that the transition 
occurs when the pressure of the quark matter equals the pressure of the hadronic matter at the same neutron chemical potential:

\begin{equation}
\mu^H_n = \mu^Q_n \quad \mbox{and} \quad p^H = p^Q, \label{EL24}
\end{equation}
where the neutron chemical potential can be written in terms of the quark ones as:
\begin{eqnarray}
\mu_d = \mu_s = \frac{1}{3} ( \mu_n + \mu_e) \nonumber \\
\mu_u = \frac{1}{3} (\mu_n - 2\mu_e) . \label{EL25}
\end{eqnarray}

\begin{figure}[htb] 
\begin{centering}
 \includegraphics[angle=270,
width=0.74\textwidth]{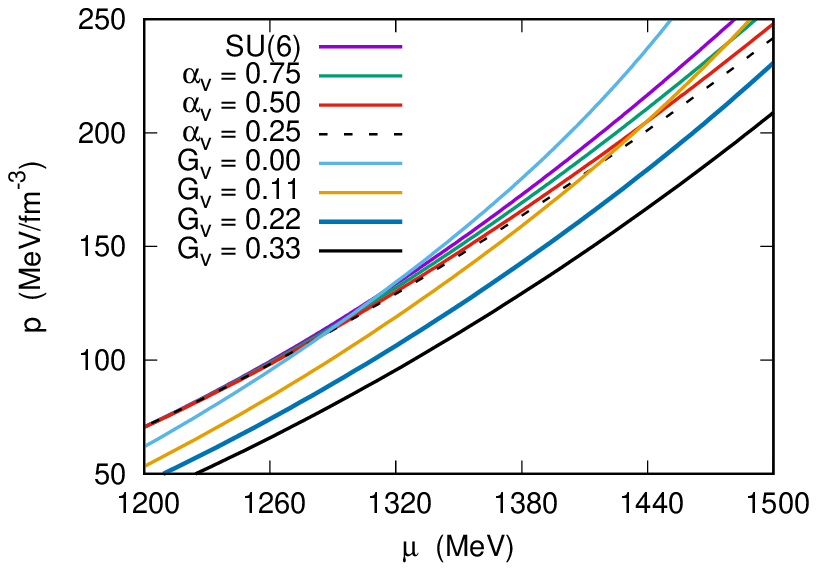}
\caption{(Color online) Pressure as a function of the neutron chemical potential for several values of $\alpha_v$ in hadronic phase and $G_V$ in quark phase.} \label{FL6}
\end{centering}
\end{figure}

\begin{center}
\begin{table}[ht]
\begin{center}
\scalebox{0.94}{
\begin{tabular}{|c|c|c|c|c|c|c|c|c|}
\hline 
 Model & $\alpha_v$ (H) &  $G_V$ (Q) & $\mu^Y$ (H)  &$\mu_n^H = \mu_n^Q$  & $p^H = p^Q$ & $\epsilon^H$(MeV$/fm^3)$ & $\epsilon^Q$(MeV$/fm^3)$ \\
 \hline
 A0 & 1.00  & 0.00$G_s$ & 1134 MeV & 1303 MeV & 123 (MeV$/fm^3)$ & 642 & 747 \\
 \hline
 B0 & 0.75  & 0.00$G_s$ & 1142 MeV & 1291 MeV & 115 (MeV$/fm^3)$ & 596 & 723 \\
 \hline
 C0 & 0.50  & 0.00$G_s$ & 1142 MeV & 1285 MeV  & 111 (MeV/$fm^3)$ & 564 & 710 \\
  \hline
 D0 & 0.25  & 0.00$G_s$ &1115 MeV & 1281 MeV  & 108 (MeV/$fm^3)$ & 542  & 701 \\
  \hline
  B1 & 0.75  & 0.11$G_s$ & 1142 MeV & 1475 MeV  & 237 (MeV/$fm^3)$ & 896  & 1159 \\
  \hline
  C1 & 0.50  & 0.11$G_s$ & 1142 MeV & 1437 MeV  & 203 (MeV/$fm^3)$ & 778  & 995 \\
  \hline
  D1 & 0.25  & 0.11$G_s$ & 1115 MeV & 1418 MeV  & 187 (MeV$/fm^3)$ & 711  & 929 \\
  \hline
\end{tabular}}
\caption{Chemical potential and pressure at phase transition for hadron (H) to quark (Q) with different values of $\alpha_v$ and $G_V$ respectively. We also show the energy density at both phases at the critical neutron chemical potential, and the chemical potential of the hyperon threshold ($\mu^Y$).}
\label{TL6}
\end{center}
\end{table}
\end{center}

We have seen that in the hadronic phase, when $\alpha_v$ has its value
reduced as it moves away from the SU(6) group,  the hyperon repulsion
increses reducing the strangeness fraction  and therefore producing
stiffer EoS and consequently, more massive neutron stars. In the same 
way, the increse of the $G_V$ value in the quark phase has the same
effect: it reduces the strangeness fraction in the core of massive
quark stars, stiffening the EoS and producing more massive quark
stars. Now we look what is the effect of a stiff/soft EoS in both
phases in hybrid stars.
To accomplish that, we plot in Fig.~\ref{FL6} the pressure as function of the neutron chemical potential, and seek for the point that satisfies the conditions presented in Eq.~(\ref{EL24}) for different values of $\alpha_v$ and $G_V$.

We see that reducing  $\alpha_v$ in the hadron phase increases the
pressure, so it induces the hadron quark phase transition at early densities when
compared with the SU(6) symmetry parametrization. On other hand, the
vector channel in the quark phase increases the pressure as well.
 With the inclusion of the vector channels, the pressure becomes
higher in the quark phase than in the hadronic one. For $G_V =0.22G_s$ the quark phase is already energetically unfavorable,  suppressing
the  phase transition. We see that only for low or zero values of $G_V$ the 
quark-hadron phase transition is possible. For larger values of $G_V$
Eq.~(\ref{EL24}) is not satisfied.  Similar behavior has already been
noticed in \cite{Deb2014}.
The phase transition pressure and the critical neutron chemical potential values are presented in Tab.~\ref{TL6}.

\begin{figure*}[ht]
\begin{tabular}{cc}
\includegraphics[width=5.6cm,height=7.0cm,angle=270]{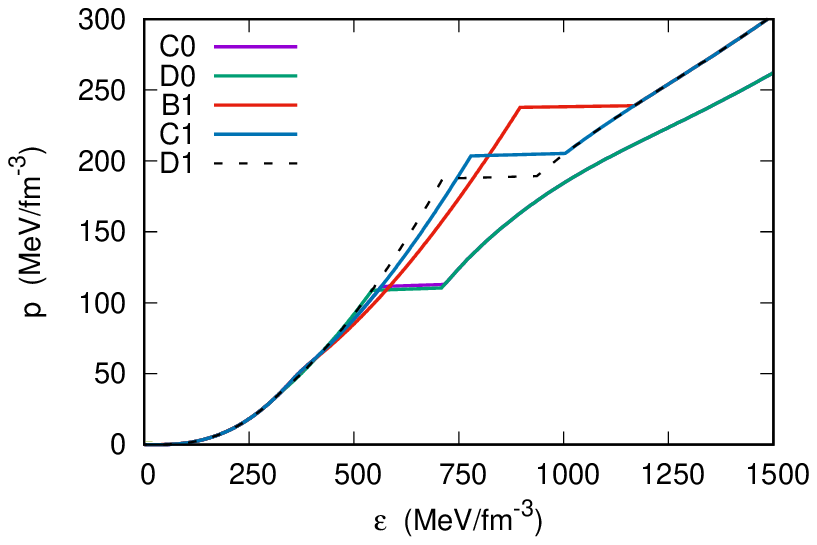} &
\includegraphics[width=5.6cm,height=7.0cm,angle=270]{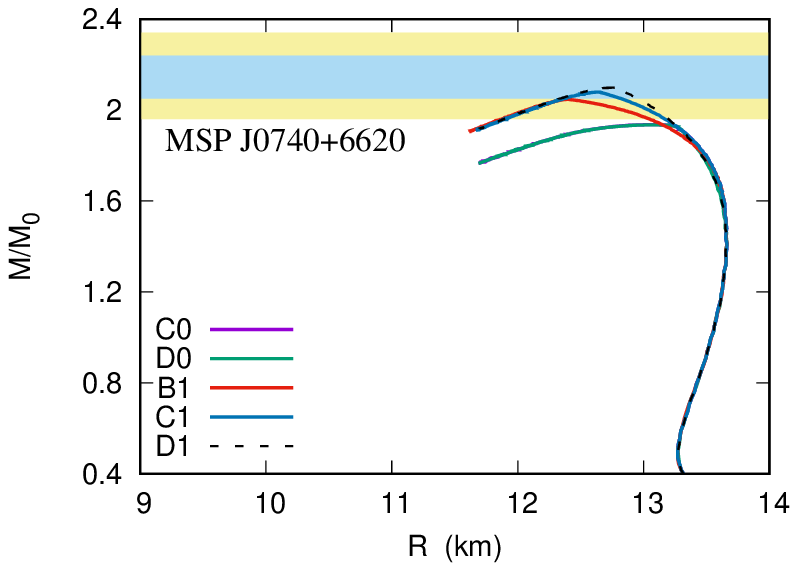} \\
\end{tabular}
\caption{(Color online) EoS and mass-radius relation for models that allow a construction of a hybrid star.} \label{FL7}
\end{figure*} 

\begin{table}[ht]
\begin{center}
\begin{tabular}{|c|c|c|c||c|}
\hline
 Model & $M/M_\odot$ & $ R (km)$ & $n_c$ ($fm^{-3}$) & $M_{min}/M_\odot$   \\
\hline
 A0     & 1.92  & 12.94   & 0.74   & 1.91  \\
\hline
 B0         & 1.93  & 13.00  &  0.73   & 1.91 \\
 \hline
C0        & 1.94 & 13.07  & 0.71   & 1.92  \\
\hline
D0        & 1.94 & 13.07  & 0.71   & 1.92  \\
\hline
B1        & 2.05 & 12.40  & 0.78   & 2.05  \\
\hline
C1        & 2.08 & 12.64  & 0.80   & 2.08  \\
\hline
D1        & 2.10 & 12.75  & 0.83   & 2.09  \\
\hline
\end{tabular}
 \caption{Hybrid star properties indicating that  the allowed mass
   values for a quark core lie within a very narrow band. }\label{TL7}
\end{center}
 \end{table}
 
 As can be seen, for $G_V$ = 0.00, all hadronic EoS allow phase transitions.
For $G_V = 0.11G_s$, all but SU(6) allow phase transitions. The lowest critical
neutron chemical potential to allow a phase transition is 1281 MeV, at the 
stiffer hadronic ($\alpha_v$ = 0.25) EoS and the softer quark EoS
($G_v = 0.00)$. On the other hand, the higher 
critical neutron chemical potential which still allows a phase transition
is 1475 MeV,  with $\alpha_v = 0.75$ and $G_V = 0.11G_s$.

Another important issue is related to the hyperon puzzle. Can the 
quark-hadron phase transition suppress the hyperon threshold 
and solve the hyperon puzzle? As can be seen from Tab.~\ref{TL6}
the answer is negative. For all values of $\alpha_v$, the $\Lambda^0$
onset (the first hyperon to appear in all cases) occurs at lower
chemical potential ($\mu^Y$) than the critical chemical potential. Our study indicates that the hyperon puzzle is stil open
even when we are dealing with quark-hadron phase transition.

In order to not saturate
the figure, we plot five of the seven possible hybrid star EoS and
the mass-radius relation in Fig.~\ref{FL7}. The main properties
of the hybrid stars are presented in Tab.~\ref{TL7}.

As we can see in this work, where all hadronic EoS are derived from
the GM1 parametrization~\cite{Glen2}, and the quark EoS from HK parametrization
~\cite{Kun2},  it is the quark EoS (and not the hadronic one) that
plays the more important role in producing different maximum mass
values. Although it is model dependent, we believe that qualitatively
these features will be maintained with other parameter choices.
Fixing $G_V = 0.00$ and varying $\alpha_v$ from 1.00 to 0.25
we produce maximum masses in the range of $1.92M_\odot$ to $1.94M_\odot$.
However, when we fix $G_V = 0.11G_s$ the maximum mass now varies
from $2.05M_\odot$ to $2.10M_\odot$. This also indicates that
hybrid stars are possible in the context of the massive MSP J0740+6620,
only if we consider a vector channel in the NJL model.
 
 Another point  to investigate is  the
$M_{min}$, which is the  minimum mass  star that supports a quark core, i.e., 
stars with masses below  $M_{min}$ are purely hadronic. We see that,
although the existence of a hybrid star is possible, it is very
unlikely, because there is just a narrow mass range that 
  supports a quark core, as already pointed in ref.~\cite{Lopes2020}.
   Indeed in some cases, the hybrid star branch is lower than  0.01$M_\odot$, and in other is around 0.02$M_\odot$. 
Our results is in agreement with those found in ref.~\cite{Sandoval},
although the authors  consider only one hyperon parametrization; also ref.~\cite{Sandoval} uses constant speed of sound parametrization to determine the phase transition while we use critical neutron chemical potential - $\mu^Q =\mu^H$ at $p^Q = p^H$.  The small branch  of the hybrid stars is much  lower than the experimental  uncertainty on the mass of the
MSP 0740+6620. Hence, its true nature is still an open subject,
although  it is unlikely that it is a quark star. Also, hybrid stars
always have  low central densities when compared with both, quark and
hadronic stars.  Of course, these results are model dependent and other hadronic   and quark models can give different quantitative results.

\subsection{The speed of sound and the mass and radius of the
quark core}

Now we confront our results in the light of a very recent study
about the nature of massive stars. In ref.~\cite{Annala}, the authors
suggest that massive neutron stars have sizable quark core. Moreover, 
the authors link the size and mass of quark core with the
quark matter speed of sound and conclude that the lower the speed
of sound of the quark matter, the higher the mass and radius of the 
quark core. If the square of the speed of sound does not strongly violate
the conformal bound - $v_s^2 < 1/3$  - a quark core 
of mass 0.8$M_\odot$ and radius around 7 km can be obtained.
We start by defining the square of the speed of sound as~\cite{Glen}:

\begin{equation}
 v_s^2 = \bigg | \frac{\partial p}{\partial \epsilon} \bigg | . \label{SS}
\end{equation}

\begin{figure}[htb] 
\begin{centering}
 \includegraphics[angle=270,
width=0.74\textwidth]{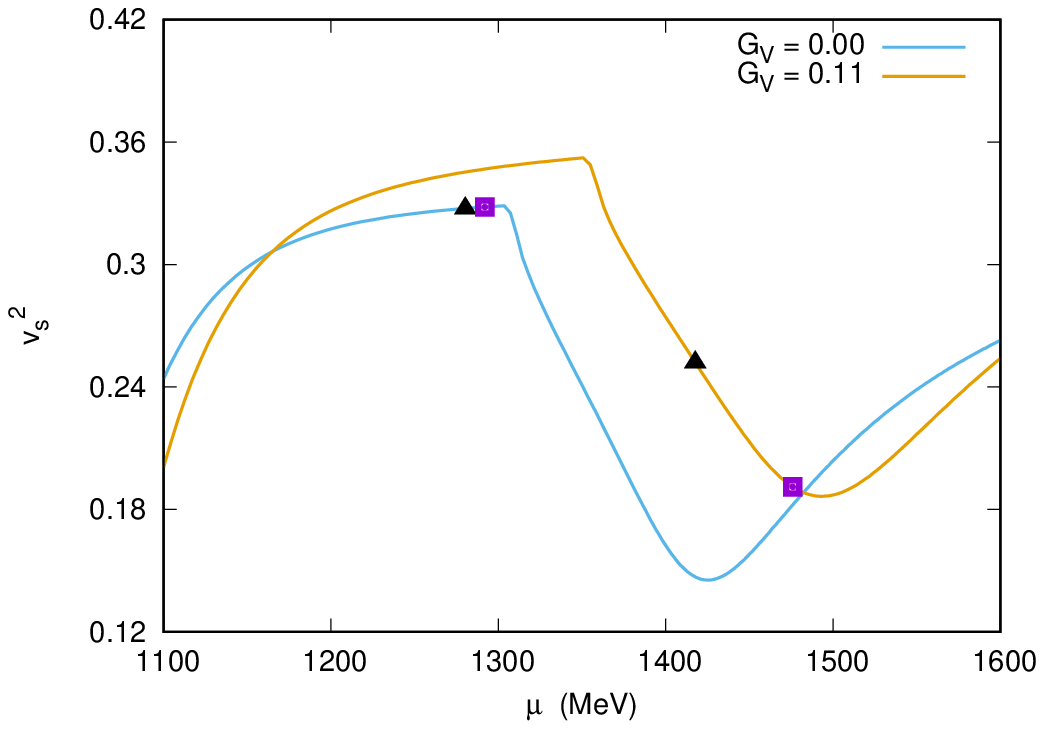}
\caption{(Color online) Speed of sound for NJL models. The black
  triangle (purple square) indicates the critical chemical potential
  for $\alpha_V$ equals to 0.25 (0.75).} \label{F8}
\end{centering}
\end{figure}

Now we plot in Fig.~\ref{F8} the speed of sound for the NJL model with $G_V = 0.0G_s$ and $G_V = 0.11G_s$ in function of the chemical potential. We also display the critical chemical potential in forms of black triangles and
purple squares for the hadronic phase within $\alpha_V$ equals to 0.25
and 0.75 respectively. For $G_V = 0.00G_s$ the speed of sound for all
analysed densities is within the conformal bound - $v_s^2 < 1/3$, while
for $G_V = 0.11G_s$ there is a small violation around 1300 MeV to 1400 MeV.
However the speed of sound at the critical potential is smaller for 
$G_V = 0.11G_s$. This is due to a peculiar behaviour of the speed of
sound in NJL (also pointed in ref.~\cite{Sandoval}), where at
certain density the speed of the sound starts to decreases and then increases
again.

To finish our analyses we estimate the mass and size of the quark
core in the maximally massive hybrid star of each model presented
in Tab.~\ref{TL7}. To accomplish that we solve the TOV equations 
for the quark EoS from the density correspondent to the critical 
pressure displayed in Tab.~\ref{TL6} up to the density at the 
maximum mass shown in Tab~\ref{TL7}. The results are presented
in absolute and relative values in  Tab~\ref{TL8}.

\begin{center}
\begin{table}[ht]
\begin{center}
\scalebox{0.99}{
\begin{tabular}{|c|c|c|c|c|c|c|c|c|}
\hline 
 Model &A0 &  B0 & C0  & D0  & B1 & C1 & D1 \\
 \hline
 $M_Q/M_\odot$  & 0.052  & 0.082 & 0.064 & 0.079 & 0.006 & 0.008 & 0.014 \\
 \hline
 R$_Q$  (km) & 2.61  & 3.06 & 2.85 & 3.05 & 1.10 & 1.30 & 1.57 \\
 \hline
 $\%$ $M_Q/M_{max}$  & 2.7$\%$  & 4.2$\%$ & 3.3$\%$  & 4.0$\%$  & 0.3$\%$  & 0.4$\%$  & 0.7$\%$ \\
 \hline
 $\%$  R$_Q$/R$_{total}$ & 20$\%$   & 24$\%$ & 22$\%$  & 23$\%$  & 8.9$\%$  & 10$\%$  & 12$\%$  \\
 \hline
\end{tabular}}
\caption{Masses and radii of the quark core and their proportional
  contribution for the maximally massive star within different hybrid
  stars models defined in Tab.~\ref{TL6}.}
\label{TL8}
\end{center}
\end{table}
\end{center}  
 
 We see that  as in the case of the hybrid branch, it is the quark EoS (over
 the hadronic one) that produces more massive and larger radii 
 quark cores. For $G_V = 0.00G_s$ the most massive hybrid star 
 can bear a quark core mass around 3$\%$ to 4$\%$, while for 
 $G_V = 0.11G_S$ this contribution barely reaches 0.5$\%$. In the same 
 way, for $G_V = 0.00G_s$ the size of the quark core can reach almost
 1/4 of the star, while for $G_V = 0.11G_s$ the contribution is only
 about 10$\%$. In all cases the contribution of the quark phase is more
 signficant in the size than in the mass. Our study corroborates 
 the results of ref.~\cite{Annala}, although we found that the quark
 core plays a much more subtle role than discussed in ref.~\cite{Annala},
 where its size can reach 60$\%$ and the quark core mass reaches 40$\%$
 of the total mass. Ultimately our results are much closer
 with those presented in ref.~\cite{Sandoval}, although we use different
 criteria for the quark-hadron phase transition, as well as different 
 parametrizations.

   \section{Final Remarks}

In this work we broke the hybrid group SU(6) in favor of a more general
SU(3) flavor symmetry group to fix the meson-hyperon couplings and employ
an additional vector field, the strangeness-hidden $\phi$ meson, to
obtain massive hadronic and hybrid stars.
We also investigate the influence of an additional vector channel
in the quark phase in the context of NJL models. The main 
results are summarized below.

\begin{itemize}

\item As we move away from the SU(6) group, by reducing $\alpha_v$ we
  produce stiffer EoS, higher maximum mass and low strangeness fraction.
Although the hyperon onset is well known to soft the EoS, we are still
able to produce 2.21$M_\odot$ hyperonic stars.

\item  In the quark phase, $G_V$ also increases the pressure, stiffing
the EoS. However for realistic values of $G_V$ our maximum mass is quite
below the MSP J0740+6620. This possibly indicates that massive pulsars
are not quark stars. Although we are  able to reproduce a 2.05$M_\odot$
quark star at the price of using $G_V = G_s$, we believe this 
vector channel value is artificial and unphysical. 
 
\item Stiffening the QHD EoS by reducing $\alpha_v$ favors the quark-hadron
phase transition while stiffening the NJL EoS by increasing $G_V$ makes the phase transition more difficult. Indeed for values of $G_V$ above $0.11G_s$ the quark-hadron phase transition is completely suppressed.

\item The quark-hadron phase transition does not solve the hyperon
puzzle, as the onset of the $\Lambda^0$ hyperons happens for densities
lower than those needed for the phase transition to occur for all values 
of $\alpha_V$ and $G_V$.

\item Within our models, which preserve all the nuclear properties of
  Tab.~\ref{TL1}, as well the quark properties of Tab.~\ref{TL4}, it is the quark EoS over the hadronic one that plays a crucial role in the production of more massive hybrid stars.

\item  According to the models presented in this work, 
the hybrid star branch is very small,  indicating that only stars at
the edge of mechanical stability can be  hybrid stars. If they exist,
hybrid stars are probably very rare in the universe.

\item Quark stars can still be present in nature, if the Bodmer-Witten
  conjecture is true.  Low mass hadronic neutron stars can
collapse to form a quark star. Massive neutron stars collapse into black holes. 

\item All our hadronic and hybrid stars present a radius of 13.68 km
  for   the canonical value of 1.4$M_\odot$. Altough it is very close to
  accepted values in the literature~\cite{Hebeler,malik,PRL121}, the
  possible 11.9 km discussed in ref.~\cite{Capano2020}, adds an
  additional puzzle either to the EoS or to general relativity. Yet, our
  results are in agreement with the recent measurements coming from 
NICER~\cite{NICER1,NICER2}.
  
\item We found that the mass of the quark core in a hybrid star is always 
  small, but the radius of the quark core can occupy 1/4 of the star. Qualitatively, our results corroborate the ones coming from ref.~\cite{Annala}, but in much more subtle way. 

\item We can describe even more massive neutron stars if hyperons are not
  present. However, our results here, as well as others like
  ref.~\cite{HI} indicate that hyperons are like Thanos, inevitable.

\end{itemize}

{\bf Acknowledgments} 
This work is dedicated to Liz ``Lizuda''. 
it is a part of the project INCT-FNA Proc. No. 464898/2014-5 and it was partially supported by CNPq (Brazil) under grant 301155.2017-8~(D.P.M.).


\section*{References}





\begin{thebibliography}{99}

\bibitem{Cromartie}
H.~Cromartie et al: Nat. Astr. {\bf 4}, 72 (2020)

\bibitem{Antoniadis}
J.~Antoniadis et al: Science {\bf 340}, 1233232 (2013)

\bibitem{Negreiros_2013} Negreiros et. al., Phys.Lett. B 718 (2013) 1176.


\bibitem{Dex4}
B.~Franzon, V.~Dexheimer, S. Schramm
Mon. Not. Roy. Astr. Soc.  {\bf 456}, 2937 (2016)


\bibitem{Dex6}
R. Gomes, V. Dexheimer,  S. Schramm
Phys. Rev. D {\bf 94} 044018 (2016)

\bibitem{HI}
H.~Djapo, B.~Schaefer, J.~Wambach Phys. Rev. C  {\bf 81}, 035803 (2010)

\bibitem{Endgame}  {\it Avengers: Endgame}, Russo brothers, Marvel Studios,
Film (2019)

\bibitem{Witten}
E.~Witten,Phys. Rev. D  {\bf 30}, 272 (1984)

\bibitem{Bodmer}
A.~R.~Bodmer,Phys. Rev. D  {\bf 4}, 1601 (1971)

\bibitem{Mc}
L.~McLerran, R.~D.~Pisarski, Nucl. Phys. A  {\bf 796}, 83 (2007)

\bibitem{Annala}
E.~Annala et al., Nat. Phys. {\bf 16} 907, (2020)

\bibitem{Serot} B.~D.~Serot, Rep. Prog. Phys. \textbf{55}, 1855 (1992)

\bibitem{Glen} N.~K. ~Glendenning,
\emph{Compact Stars},Springer, New York -  Second Edition  (2000)

\bibitem{Ellis} J. Ellis, J. I. Kapusta, K. A. Olive
Nucl. Phys. B \textbf{348}, 345 (1991)

\bibitem{Lopes2013}
L.~L.~Lopes, D.P.~Menezes, Phys. Rev. C  {\bf 89}, 025805 (2014)

\bibitem{Rafa2005}
R.~Cavagnoli, D.P.~Menezes, Braz. J. Phys.  {\bf 35}, 869 (2005)

\bibitem{Weiss2}
S. Weissenborn, D. Chatterjee, and J. Schaffner-Bielich, Phys.Rev. C 85, 065802 (2012) 


\bibitem{Lopes2020}
L.~Lopes, D.~Menezes Eur.Phys.J. {\bf 56}, 122 (2020) 


\bibitem{Nambu}
Y.~Nambu, G.~Jona-Lasinio, Phys. Rev.  {\bf 122}, 345 (1961)


\bibitem{Lopes2016}
L.~L.~Lopes, D.P.~Menezes, Eur. Phys. J. A  {\bf 52}, 17 (2016)

\bibitem{Sugano2014} J.~Sugano et al.,
Phys. Rev. D  \textbf{90}, 037901  (2014)


\bibitem{Rafa} R.~Cavagnoli, D.~P. Menezes, C.~Providencia, Phys. Rev. 
C \textbf{84}, 065810 (2011).

\bibitem{Lopes2014}
L.~L.~Lopes, D.P.~Menezes, Braz. J. Phys.  {\bf 44}, 744 (2014)

\bibitem{Tsang}
M.~B.~Tsang et al, Phys. Rev. C {\bf 86}, 015803 (2012) 

\bibitem{Micaela2017} M. Oertel et al, Rev. Mod. Phys. {\bf89}, 015007 (2017).

\bibitem{Dutra2014} M. Dutra et al,
 Phys. Rev. C {\bf 90}, 055203 (2014).
 
\bibitem{Lattimer2014} Lattimer \& Steiner, Eur. Phys. J. A {\bf 50}, 40 (2014) 
 
\bibitem{Pais2016} H.~Pais \& C.~Providencia, Phys.Rev. C {\bf 94}, 015808 (2016) 


\bibitem{Dex2019} V.~Dexheimer et al, J.Phys. G {\bf 46}, 034002 (2019)

\bibitem{Prov2019} Providencia et al, Front. Astron. Space Sci., {\bf 26}   March 2019

 \bibitem{Boguta} J. Boguta and A.R. Bodmer,
 Nucl. Phys. A \textbf{292}, 413 (1977).

\bibitem{Glen2}N.~K.~Glendenning,  S.~A.~Moszkowski,
Phys. Rev. Lett.  {\bf 67}, 2414 -  (1991).

\bibitem{Stone} J. Stone, N. Stone  S.~Moszkowski,
Phys. Rev. C  {\bf 89}, 044316 -  (2014).


 
\bibitem{james_low}  James R. Torres, Francesca Gulminelli and
Debora P Menezes, PRC 93, 024306 (2016).

\bibitem{james_high} James R. Torres, Francesca Gulminelli and Debora
P. Menezes, PRC 95, 025201 (2017).

 
 \bibitem{Swart63} J.~J.~Swart,
Rev. Mod. Phys. \textbf{35}, 916 (1963)
 

\bibitem{Pais} A.~Pais,
Rev. Mod. Phys. \textbf{38}, 215 (1966)



 \bibitem{Sakurai60} J.~J.~Sakurai,
Ann. Phys. \textbf{11}, 1 (1960)


 \bibitem{Greiner2}
W.~Greiner, L.~Neise, H.~Stocker,
\emph{Thermodynamics and Statistical Mechanics}, 
Springer, New York,  (1995)


\bibitem{Lopes2012}
L.~L.~Lopes, D.P.~Menezes, Braz. J. Phys.  {\bf 42}, 428 (2012)


\bibitem{TOV}
 R.C. Tolman, Phys. Rev. 55, 364 (1939);
J.~R.~Oppenheimer, G.~M.~Volkoff, Phys. Rev. \textbf{33}, 374 (1939).

\bibitem{17km} C.~Wynn et al.,
Mont. Not. Roy. Astron. Soc. {\bf 375}, 821 (2007)


\bibitem{Hebeler} K.~Hebeler et al,
Phys. Rev. Lett. \textbf{105}, 161102 (2010)

\bibitem{Lattimer2013} J.M. Lattimer and A.W. Steiner,
Astrophys. J.  \textbf{784}, 123 (2014)


\bibitem{PRL121} B.~Abbott et al,
Phys. Rev. Lett. \textbf{121}, 161101 (2018)

\bibitem{malik} Tuhin Malik, N. Alam, M. Fortin, C. Provid\^encia, B. K. Agrawal, T. K. Jha, 
Bharat Kumar, and S. K. Patra Phys. Rev. C. {\bf 98}, 035804 (2018).

\bibitem{Capano2020}
C.~Capano et al: Nat. Astr. 
(2020)

\bibitem{clesio2019} Cl\'esio E. Mota, Luis C. N. Santos,
Guilherme Grams, Franciele M. da Silva and D\'ebora P. Menezes,
Phys. Rev. D 100, 024043 (2019)

\bibitem{Lopes2018} L. Lopes and D. P. Menezes,
J. Cosm. Astrop. Phys.  \textbf{05}, 038  (2018)

\bibitem{NICER1} T. E. Riley et al.,
Astrophys. J. Lett. {\bf 887}, L21 (2019)

\bibitem{NICER2} M. C. Miller et al.,
Astrophys. J. Lett. {\bf 887}, L24 (2019)

\bibitem{PRL120}F. Fattoyev, J. Piekarewicz, and C. Horowitz,
Phys. Rev. Lett. {\bf 120} 172702 (2018)


\bibitem{DebPRC2014} D. P. Menezes et al.,
Phys. Rev. C  \textbf{89}, 055207  (2014)

\bibitem{Kun1} T~Hatsuda,  T.~Kunihiro,
Phys. Lett. B  \textbf{198}, 126  (1987)

\bibitem{Kun2} T~Hatsuda,  T.~Kunihiro,
Phys. Rep.  \textbf{247}, 221  (1994)

\bibitem{kit2002} M.~Kitazawa et al.,
Prog. Theor. Phys. \textbf{108}, 5  (2002)

\bibitem{klahn} 
T.~Klahn, T.~Fischer, Astrophys.J. {\bf 810}, 134 (2015)


\bibitem{MB} R. Denke, M.~B.~Pinto,
Phys. Rev. D  \textbf{88}, 056008  (2013)

\bibitem{Hana2001} M.~Hanauske et al.,
Phys. Rev. D  \textbf{64}, 043005  (2001)


\bibitem{Rhabi} Rhabi et al, J.Phys. G36,  115204 (2009)


\bibitem{DebJCAP2019}
D.P.~Menezes et al., J. Cosm. Astrop. Phys. {\bf 01}, 024 (2019)



\bibitem{Shao} G.~Y.~Shao et al.,
Phys. Rev. D  \textbf{85}, 114017  (2012)

\bibitem{Contrera2014} G.~A.~Contrera, A.~G.~Grunfeld, D.~B.~Blaschke.,
Phys. Part. Nucl. Lett.  \textbf{11}, 4  (2014)




\bibitem{Hell2011} K.~Kashiwa, T.~Hell, W.~Weise,
Phys. Rev. D  \textbf{84}, 056010  (2011)



\bibitem{Melrose} D. P. Menezes, D. B. Melrose, C. Provid\^encia, and
  K. Wu, Phys. Rev. C 73, 025806 (2006).

\bibitem{hybrid} D. P. Menezes and C. Provid\^encia, Phys. Rev. C 68, 035804 (2003).

\bibitem{Beth} H. Beth, G. E. Brown, J.~Cooperstein
Nucl. Phys. A \textbf{462}, 791 (1987)

\bibitem{Serot2} B. Serot, H.~Uechi
Ann. Phys. A \textbf{179}, 272 (1987)

\bibitem{Chiba1} T.~Maruyama et al.,
Phys. Rev. D  \textbf{76}, 123015  (2007)

\bibitem{Chiba2} T.~Maruyama et al.,
Phys. Lett. B  \textbf{659}, 192  (2007)

\bibitem{Paoli}
M.~Paoli, D.P.~Menezes, Eur. Phys. J. A  {\bf 46}, 413 (2010)


\bibitem{Deb2014} D\'ebora P. Menezes, et al.,
Phys. Rev. C {\bf 89}, 055207 (2014).

 \bibitem{Sandoval} I. Sandoval, et al.,
Phys. Rev. C {\bf 93}, 045812 (2016).                                         
 





\end{thebibliography}
\end{document}